\documentclass[twocolumn,showpacs,preprintnumbers,amsmath,amssymb]{revtex4}
\usepackage{graphicx}
\usepackage{dcolumn}
\usepackage{bm}
\newcommand{\be}{\begin{eqnarray}}
\newcommand{\ee}{\end{eqnarray}}

\begin{document}
\title{Localization problem of the quasiperiodic system 
with the spin orbit interaction}
\author{Mahito Kohmoto and Daijiro Tobe}
\affiliation{ Institute for Solid State Physics, University of Tokyo\\
5-1-5 Kashiwanoha, Kashiwa, Chiba 277-8581, Japan}
\begin{abstract}
 We study one dimensional quasiperiodic system 
obtained from the tight-binding model on the square lattice 
in a uniform magnetic field with the spin orbit interaction. 
The phase diagram with respect to the Harper coupling  
and the Rashba coupling are proposed from a number of 
numerical studies including  a  multifractal analysis.  
There are four phases, I, II, III, and IV in this order 
from weak to strong Harper coupling. 
In the weak coupling phase I all the wave functions are extended, 
in the intermediate coupling phases II and III mobility edges exist,
and  accordingly both localized and extended wave functions exist, 
and in the strong Harper coupling phase IV all the wave functions are localized. 
Phase I and Phase IV are related by the duality, 
and phases II and III are related by the duality, as well. 
A localized wave function is related to an extended wave function by the duality,
and vice versa. The boundary between phases II and III 
is the self-dual line on which all the wave functions are critical.

In the present model the duality does not lead to pure spectra in contrast to the case of Harper equation.
\end{abstract}
\maketitle

\section{introduction}
The quasiperiodic systems 
are intermediate between  periodic and random.
Electronic wave functions are always extended in periodic systems. 
In one dimension, wave functions are always localized in  random systems. 
The one-dimentional quasiperiodic systems show various 
interesting wave functions and energy spectral characters (for a review,
see \cite{HiramotoKohmoto}).

Consider the  one-dimensional tight-binding model:
\begin{equation}
 -\psi_{n+1}-\psi_{n-1}+ V(\omega n)\psi_n=E\psi_n, 
\end{equation}
where $V(x)$ is a periodic function with period $1$, {\it i. e.} $V(x+1)=V(x)$. When $\omega$ is an irrational number, this becomes quaisperiodic.
In particular, the model defined by 
\begin{equation}
 V(x)=2\lambda \cos(2\pi x),
 \label{lambdaCos} 
\end{equation}
 is known as Harper equation. It has  Aubry duality \cite{AubryAndre} which relates the region with $\lambda>1$ and the  region with  $\lambda<1$. Thus $\lambda=1$ is the self-dual point. From this duality and Thouless's formula for Lyapunov exponent \cite{Thouless1}, it has been argued that  all the wave functions are extended for $\lambda<1$, while  all the wave functions are localized for $\lambda>1$. At $\lambda=1$, all the wave functions are critical\cite{kohmoto83}. Thus  extended, localized, and critical
wave functions do not coexist in the energy spectrum  
 as it is pure at each value of $\lambda$;
the spectrum for $\lambda < 1$ is purely absolutely continuous, 
it is purely singular continuous for $\lambda=1$, and  it is purely 
dense point for $\lambda > 1$. 

A pure spectrum, however, is not a general feature of the
one-dimensional quasiperiodic systems. It is known that the model in Ref. 4 which has
\begin{equation}
 V(x)=\lambda\tanh(\mu\cos{(2\pi x)})/\tanh{\mu}, 
\end{equation}
\cite{HiramotoKohmoto1}
and the genenalized Harper model \cite{HiramotoKohmoto2}
in which $V(x)$ has two periods:
\begin{equation}
 V(x)=\lambda_1\cos(2\pi x)+\lambda_m\cos(2m\pi x), 
\end{equation}
have non-pure spectra and mobility edge structures. 

If we choose
\begin{equation}
 V(x)=
\begin{cases}
 -\lambda,& m-\sigma_G<x\le m\\
 +\lambda,&  m<x\le m+1-\sigma_G,
\end{cases}
\end{equation}
where $\omega=\sigma_G=(\sqrt{5}-1)/2$ (the inverse of the
golden mean) and $m$ is an arbitraly integer, the potential follows the Fibonacci sequence which is constructed recursively as
\be
\{M_{\ell + 1}\} = \{M_{ \ell - 1}\}\{M_{\ell}\},
\ee
with
$\{M_{0}\} = B, \{M_{1}\} = A$. Thus $\{M_{2}\} = BA, \{M_{3}\} = ABA, \{M_{4}\} = BAABA, \cdots$.
This is called the Fibonacci model\cite{KKT,Ostlund}. From the renormalization group study it is shown that the energy spectrum is purely singular continuous and all the wave functions are critical\cite{KO,KB}. This property is proven rigorously\cite{Suto}.

The spin Hall effect is induced by spin orbit coupling in which spin current is in the transverse direction to an applied electric field in the absence of a magnetic field \cite{rashba}. This phenomenon has been attracting much interest because of rich physics and potential industrial applications. This is in contrast to the quantum Hall effect in which an applied magnetic field breaks time reversal symmetry. But the theory of quantum Hall effect plays an important role in some aspects \cite{tknn,ann,kane1,kane2}.

Some of the basic models of the spin Hall effect involve the Rashba spin orbit coupling.
It can change
properties of electron systems essentially.
For example, the level-avoiding gaps for single electron system
with disorder on a cylinder shows a new kind of universal
behavior in the presence of the Rashba spin orbit coupling.
The distribution shows an exponential decay rather than the Gaussian
one, which can not be explained by the Gaussian unitary ensemble \cite{tobe}.

In this paper the tight binding model on the square lattice in a magnetic field is considered. The Rashba spin orbit interaction is also included. The model is reduced to $1d$ and it is quasiperiodic when magnetic flux per plaquette is irrational. It is a generalization of Harper equation and we study the properties of the energy spectrum and the wave functions  of this $1d$ quasiperiodic system. 
We find four phases: the weak coupling phase I, the  strong coupling phase IV, and the intermediate coupling phases II, III (see Fig.~\ref{FigPD}). All the wave functions are extended  in the weak coupling phase I, and all the wave functions are localized  in the strong coupling phase IV. In contrast to Harper equation localized and extended wave functions coexist, and accordingly mobility edges appear in the intermediate coupling phases II and III. The boundary of these phases is the self-dual line where all the wave functions are critical, and the spectrum is singular continuous. The self-dual line is a critical point of a phase transition between a localized wave function and a extended wave function.

The Rashba spin orbit interaction does not break Aubry duality, but introduces a coupling between two Harper equations (one for spin up and the other for spin down). Our result suggests that an extended state in phase I  could be either localized or extended in phases II and III. This is also true for a localized state in phase IV. Thus the spin orbit coupling enhances tendency toward neither localization nor delocalization. This suggests that the existence of mixed spectra is a rather common property of  the $1d $ quasiperiodic system. The duality itself does not guarantee a pure spectrum. 

The organization of this paper is as follows: 
in Sec. \ref{hamiltonian} we introduce
two dimensional elecrons with the Rashba spin orbit interaction
on the square lattice in a uniform magnetic field and obtain the
one-dimensional quasiperiodic system. The duality which is a generalization of  Aubry duality of Harper equation is discussed. The proposed phase diagram is shown and explained in Sec. \ref{PhaseDiagram}.
In Sec. \ref{Scaling} the scaling approach is explained. The analysis of the total band widths is presented which supports the proposed phase diagram. In Sec. \ref{Multifractal} the multifractal method for an energy spectrum and wave functions is introduced. The scheme for identifying phases with absolutely continuous, singular continuous, dense point, and mixed energy spectra are given. In Sec. \ref{WF} examples of extended, localized, and critical wave functions are shown. 

\section{Hamiltonian and Duality}
\label{hamiltonian}
\subsection{Electron hopping on the square lattice in a magnetic field with the spin orbit interaction}
Consider tight binding electrons on the square lattice in a
perpendicular magnetic field with the Rashba spin orbit interaction \cite{rashba}. 
The problem involves a Hamiltonian: 
\be
 H &=& -\sum_{n,m}(t_x\hat{c}^\dagger_{n+1,m}\hat{c}_{n,m}e^{i\theta^x_{n,m}} +
t_y\hat{c}^\dagger_{n,m+1}\hat{c}_{n,m}e^{i\theta^y_{n,m}}) \nonumber \\
&+& i\lambda_{R}\sum_{n,m} ( \hat{c}^\dagger_{n,m+1}\sigma_x\hat{c}_{n,m}e^{i\theta^y_{n,m}}
-\hat{c}^\dagger_{n+1,m}\sigma_y\hat{c}_{n,m}e^{i\theta^x_{n,m}} ) \nonumber \\
&+&  {\rm H.c.},
\label{ham-1}
\ee
where $\hat{c}_{n, m}= {^t}(c_{n, m\uparrow}, c_{n, m\downarrow})$ and $\hat{c}_{n, m}^\dagger = (c_{n, m\uparrow}^\dagger,c_{n, m\downarrow}^\dagger)$ ($t$ is a transpose).  The transfer integrals in $x$- and $y$-directions are given by $t_x$ and $t_y$, respectively, and $\lambda_{R}$ is the Rashba spin orbit coupling. 
When $\phi=0$, Eq. (\ref{ham-1}) reduces to the Rashba
model on the square lattice.

We take the Landau gauge:
\begin{equation}
 \theta_{n,m}^x=0, \; \theta_{n,m}^y=2\pi\phi n,
\label{LandauGauge1 }
\end{equation}
and flux per plaquette is given by $2\pi \phi$. 
 With this choice of gauge the Hamiltonian has translational symmetry in $y$-direction and one can write 
\be
 |\psi\rangle =\sum_{n,m,\sigma=(\uparrow,\downarrow)}
e^{ik_y m}\psi_{n,\sigma}c^\dagger_{n,\sigma}|0\rangle,
\ee
then $H |\psi\rangle=E |\psi\rangle$ leads to
\begin{align}
&-t_x(\psi_{n+1,\uparrow} + \psi_{n-1,\uparrow})
-2t_y \cos{(2\pi\phi n -k_y)}\psi_{n,\uparrow} \nonumber \\
&+\lambda_{R}\left\{\psi_{n+1,\downarrow} - \psi_{n-1,\downarrow}
-2\sin{(2\pi \phi n - k_y)}\psi_{n,\downarrow}
\right\} =E\psi_{n,\uparrow}, \nonumber \\
&-t_x(\psi_{n+1,\downarrow} + \psi_{n-1,\downarrow})
-2t_y\cos{(2 \pi\phi n - k_y)}\psi_{n,\downarrow} \nonumber \\
&+\lambda_{R}\left\{ - \psi_{n+1,\uparrow} + \psi_{n-1,\uparrow}
-2\sin{( 2\pi \phi n - k_y)}\psi_{n,\uparrow}\right\} =E\psi_{n,\downarrow},
\label{tb1}
\end{align}

Let us make a gauge transformation by exchanging $x$- and $y$-directions. The new gauge is
\begin{equation}
 \theta_{n,m}^x=2\pi\phi m, \; \theta_{n,m}^y=0,
 \label{LandauGauge2} 
\end{equation}
then we have
\begin{align}
 &-t_y(f_{m+1,\uparrow} + f_{m-1,\uparrow}) -2t_x\cos(2\pi\phi m - k_x)f_{m,\uparrow} \nonumber \\
&+\lambda_{R}\left\{ f_{m+1,\downarrow} - f_{m-1,\downarrow}-2\sin(2\pi\phi m - k_x)f_{m,\downarrow} \right\}
=Ef_{m,\uparrow}, \nonumber \\
 &-t_y(f_{m+1,\downarrow} + f_{m-1,\downarrow}) - 2t_x\cos( 2\pi\phi m - k_x) f_{m,\downarrow}\nonumber \\
&+\lambda_{R}\left\{ - f_{m+1,\uparrow} + f_{m-1,\uparrow}-2\sin(2\pi\phi m - k_x)f_{m,\uparrow} \right\}
=Ef_{m,\downarrow},
\label{tb2}
\end{align}
where,
\be
 |\psi\rangle =\sum_{n,m,\sigma=(\uparrow,\downarrow)}
e^{ ik_x n}\psi_{m,\sigma}c^\dagger_{m,\sigma}|0\rangle,
\ee

The $1d$ wave functions in Eqs. (\ref{tb1}) and (\ref{tb2}) are related by a Fourier transformation,
\be
 \psi_{n,\sigma}=\alpha(\sigma) 
e^{ -ik_x n}\sum_{m}e^{i(2\pi\phi n - k_y)m} f_{m,\sigma},
\label{FT}
\ee
where
\be
\alpha(\uparrow)=1
,\quad
\alpha(\downarrow)= i.
\ee

\subsection{Quasiperiodic system in one dimension and the duality}
Equation (\ref{tb1}) represents a tight binding model in $1d$. When $\phi$ is a rational number,  $\phi = p/q$, where $p$ and $q$ are mutually prime integers, it is periodic with period $q$. Thus Bloch's theorem is valid and the wave functions are of Bloch form and extended. The energy spectrum consists of $2q$ bands.

When $\phi$ is an irrational number, which could be considered as $p \rightarrow \infty$ and $q \rightarrow \infty$, Bloch's theorem cannot be applied. Thus it is not certain whether the wave functions are extended or not. In addition, since the number of bands $2q$ is infinite, the energy spectrum is expected to be a Cantor set. This is the localization problem we study.

In order to construct a phase diagram the duality plays a key role. Compare Eqs. (\ref{tb1}) and (\ref{tb2}) and one notices that they have the same form except that  $t_x$ and $t_y$, as well as $k_x$ and $k_y$,  are exchanged. The wave functions are related by the Fourier transformation (\ref{FT}). Let us introduce a parameter,
\be
\lambda_H = \frac{t_y}{t_x},
\label{lambda}
\ee
then the system is self-dual if $\lambda_H = 1$. A wave function for $\lambda_H = a$ is related to the one for $\lambda_H = 1/a$ by the Fourier transformation (\ref{FT}). Thus, if a wave function is exponentially localized for $\lambda_H = a$, for example,  the corresponding wave function for $\lambda_H = 1/a$ is extended.

If $\lambda_R=0$, there is no coupling between electrons with opposite spins and the system is reduced to Harper equation. In this case the duality  is called Aubry duality\cite{AubryAndre}. Thus the duality in the present system is a generalization of Aubry duality.

\section{Phase Diagram}
\label{PhaseDiagram}
If a energy spectrum is pure, there are three possibilities: absolutely continuous, dense point, and singular continuous. For an absolutely continuous spectrum all the wave functions are extended, for a dense point spectrum all the wave function are localized, and for a singular continuous spectrum all the wave functions are critical. Another possibility is a mixed spectrum which consists of absolutely continuous parts and dense point parts. Therefore extended wave functions and localized wave functions coexist in an energy spectrum in this case.

\begin{figure}[t]
 \includegraphics[width=70mm]{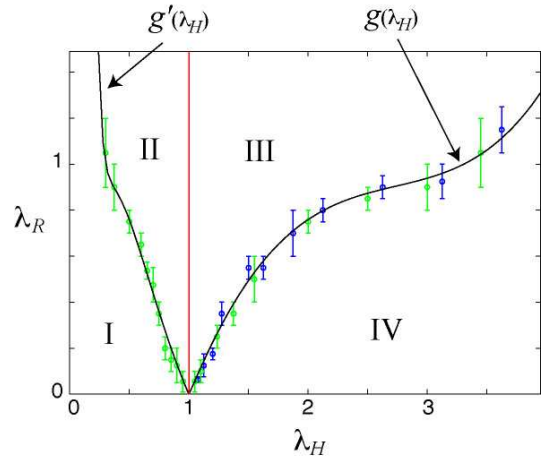} 
\caption{\label{FigPD} (Color online) Proposed phase diagram. Phase I and Phase IV are dual each other, as well as Phase II and Phase III. The system is self-dual on the red line, $\lambda_H = 1$; the energy spectrum is purely singular continuous and all the wave functions are critical. Phase I: The energy spectrum purely absolutely continuous and all the wave functions are extended. Phase II and Phase III: The energy spectrum is mixed and consists of absolutely continuous parts and dense point parts. Phase IV: The energy spectrum is purely dense point and all the wave functions are localized.}
\end{figure}

In Fig.~\ref{FigPD} we show the proposed phase diagram for the present quasiperiodic $1d$ system. In the week coupling phase I the spectrum is absolutely continuous and all the wave functions are extended. 
In the intermediate phases II and III the spectrum is mixed; the absolutely continuous parts and dense point parts coexist and they are separated by mobility edges. These two phases are related by the duality. So the absolutely continuous parts and dense point parts of the spectrum are exchanged by the duality which is discussed in Sec. \ref{hamiltonian} B. The extended wave functions and the localized wave functions are related by the transformation (\ref{FT}).

The boundary of Phase II and Phase III is the self-dual line, $\lambda_H =1$. We expect that all the wave functions are critical and the spectrum is purely singular continuous.

In the strong coupling phase IV the spectrum is dense point and all the wave functions are localized. This strong coupling phase IV and the weak coupling phase I are related by the duality. The corresponding wave functions are related by the transformation (\ref{FT}) and all the wave functions are extended in Phase I.

\section{The scaling approach}
\label{Scaling}
We analyze the energy spectrum and the wave functions by the scaling approach. 
We choose the inverse of the golden mean, $\sigma_G = (\sqrt{5} -1) /2$, for
irrational number $\phi$. Note the relation,
\be
 \lim_{\ell \rightarrow \infty}\frac{F_{\ell-1}}{F_{\ell}}=\sigma_G,
\ee
where the Fibonacci numbers $F_{\ell}$ are defined recursively by  $F_{\ell+1}=F_{\ell-1}+F_{\ell}$, with $F_0=F_1=1$. 
We take  rational approximations $\phi_\ell = F_{\ell-1}/F_{\ell}$ in which the system is periodic with period $F_{\ell}$. Then systematic studies for large $\ell$ are performed.

When $\phi_{\ell}=F_{\ell-1}/{F_{\ell}}= p/q$, 
Eq. (\ref{tb1}) is periodic with period $q$. 
From Floquet's or Bloch's  theorem
we have $\psi_{n+q}=e^{ikq}\psi_n \; (n = 1,2, \cdots, q)$. Thus, if we introduce a vector $\psi=
(\psi_{1,\uparrow},\psi_{1,\downarrow},\psi_{2,\uparrow},\psi_{2,\downarrow},\cdots,\psi_{q-1,\uparrow},
\psi_{q-1,\downarrow}, \psi_{q,\uparrow}, \psi_{q,\downarrow})$, Eq.(\ref{tb1}) is reduced to an eigenvalue problem of a  
$2q\times 2q$ matrix: 
\begin{equation}
 H_q=
\begin{pmatrix}
 A_1&B&0&\cdots& & &0 &C\\
B^{\dagger}&A_2&B&0&\cdots& & &0\\
0&B^\dagger&A_3&B&0&\cdots& &0\\
0&0&B^\dagger&A_4&B&0&\cdots&0\\
\vdots&\ddots&\ddots&\ddots&\ddots&\ddots&\ddots&\vdots\\
0& &\cdots &0 &B^\dagger &A_{q-2} &B &0\\
0& & &\cdots &0 &B^\dagger &A_{q-1} &B\\
C^\dagger& & & &\cdots &0 &B^\dagger &A_q
\end{pmatrix},
\label{num-1}
\end{equation}
where 
\begin{equation}
 A_n= -2
\begin{pmatrix}
 t_y\cos{(2\pi\phi_\ell n)} & \lambda_R\sin{(2\pi\phi_\ell n)}\\
 \lambda_R\sin{(2\pi\phi_\ell n)} & t_y\cos{(2\pi\phi_\ell n)}
\end{pmatrix},
\end{equation}
\be
 B=
\begin{pmatrix}
 -t_x &\lambda_R\\
- \lambda_R&-t_x
\end{pmatrix},
\ee
and
\be
C=
\begin{pmatrix}
 -t_x & -\lambda_R\\
\lambda_R &-t_x
\end{pmatrix}
e^{-iqk_x}.
\ee
We set $k_y=0$ for simplicity in Eq. (\ref{tb1}). The first Brillouin zone of this $1d$ system is
\be
-\frac{\pi}{q} \le k_x \le \frac{\pi}{q}.
\ee
Then, the energy spectrum has $2q = 2F_{\ell}$ bands.

\subsection{Total band width and the phase diagram}
Quantities which are useful in investigating phase transitions of quasiperiodic systems 
are sums of band widths of rational approximants  which are denoted by $B$\cite{kohmoto83,Thouless2}.
If $B$ is non-zero in the quasiperiodic limit, $\ell \rightarrow \infty$, 
the energy spectrum has absolutely continuous parts 
and the corresponding wave functions are extended.  On the other hand, wave functions are localilzed or critical when $B$ tends to zero. 

\begin{figure}[t]
 \includegraphics[width=65mm]{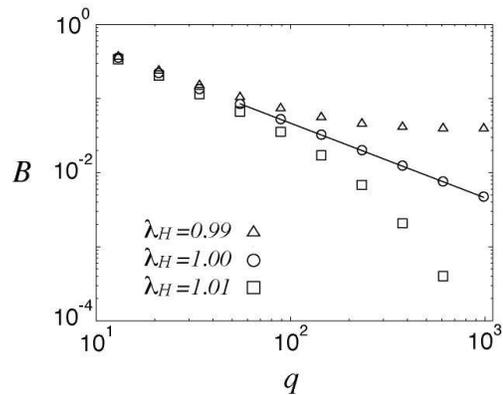}
\caption{\label{BW-harper} Total band widths of Harper equation. The limit
 $q \rightarrow\infty$ represents the quasiperiodic system.}
\end{figure}

When $\lambda_R=0$, there is no spin orbit interaction and the problem is for electrons on the square lattice in a magnetic field, namely Harper equation. As a canonical example of the scaling approach, results for $B$ in Harper system are shown in Fig.~\ref{BW-harper}. For $\lambda_H < 1$, $B$ approach a finite value for $q \rightarrow \infty$ while for $\lambda_H > 1$, $B$ tends to zero rapidly. So the energy spectrum is dense point for $\lambda_H > 1$. The result for $\lambda_H < 1$ suggests that the energy spectrum contains an absolutely continuous parts. But with the aide of the duality which connects the two regions it can be concluded that the energy spectrum is purely absolutely continuous for $\lambda_H < 1$.
For $\lambda_H=1$ the total band width behaves as 
\be
B\sim q^{-\delta}. 
\label{delta}
\ee
This behavior suggests a singular continuous spectrum and 
the exponent $\delta$ is an index which 
characterizes the criticality  and is estimated in Ref. 4 to be $1.00$. 

\begin{figure}[t]
 \includegraphics[width=65mm]{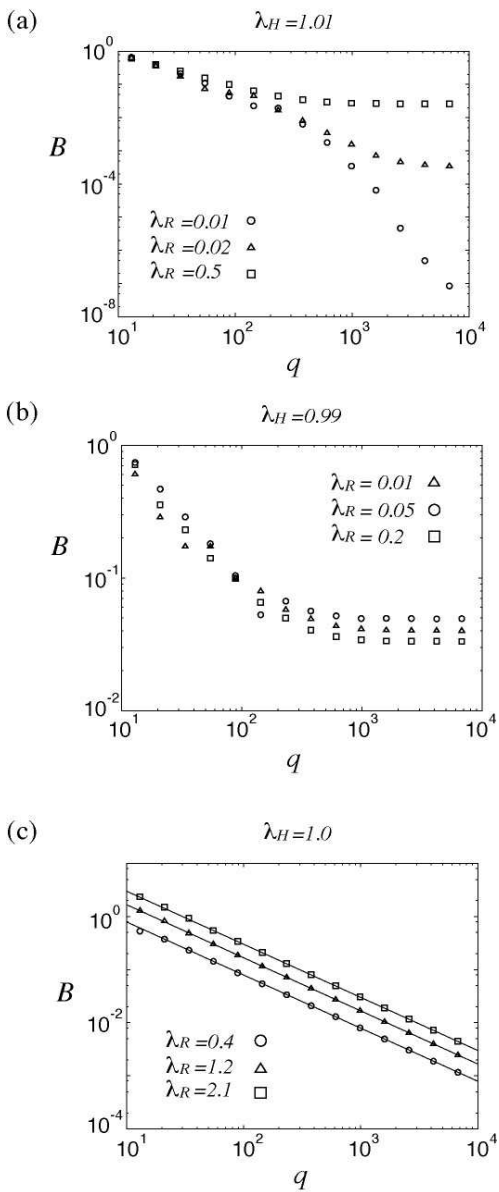}
\caption{\label{BW} Total band widths for $\lambda_R > 0$. The limit
 $q \rightarrow\infty$ represents the quasiperiodic system.}
\end{figure}

The scaling method for $B$ is applied to the present model Eq. (\ref{tb1}). In Fig.~\ref{FigPD}, the boundary between Phase III and Phase IV is denoted by $g(\lambda_H)$. Examples of data which determine $g(\lambda_H)$ is shown in Fig.~\ref{BW}(a). For $\lambda_R > g(\lambda_H)$ $B$ tends to a finite value, while for $\lambda_R < g(\lambda_H)$ $B$ tends to zero.  In Pase III there are absolutely continuous parts in the energy spectrum, but there is no absolutely continuous parts in Phase IV. 

The green error bars in Fig. \ref{FigPD} are determined by observing that, just outside the errors bars, the total band widths tend to a finite value or go to zero exponentially within the numerical accuracy.

 The boundary between Phase I and Phase II,
 $g'(\lambda_H)$, is related to $g(\lambda_H)$ by the duality as
\be
g'(\lambda_H) = g(\frac{1}{\lambda_H}),
\ee
and is shown in Fig.~\ref{FigPD}. In Phase I and Phase II $B$ always tends to a finite value as shown in Fig.~\ref{BW}(b).

The function is estimated as
\be
g(\lambda_H) & = & a(\lambda_H-1)^\beta+ b(\lambda_H-1)^{\beta-1} \nonumber \\
                          &    &  + c(\lambda_H-1)^{\beta-2},
\ee
where $a=0.14\pm 0.02$, $b=-0.72\pm 0.09$, $c= 1.34\pm 0.09$ and $\beta=3.03 \pm 0.08$. 

On line $\lambda_H=1$ the system is self-dual. The scaling analysis shows that the system is always critical. Examples of the scaling analysis are given in Fig.~\ref{BW}(c).
The scaling index defined by Eq. (\ref{delta}) is independent of $\lambda_R$ and is given by $\delta = 1.00$. 

\section{Multifractal analysis}
\label{Multifractal}
We further investigate scaling property of the energy spectrum and the wave functions 
by the multifractal analysis \cite{Halsey}  to strengthen the proposed phase diagram. 
In Ref.\cite{Kohmoto_entropy}  a general formulation of the multifractal theory 
is introduced which reformulates the
theory along the line of the  standard statistical mechanics. This method can be applied to both energy spectra and wave  functions of quaisiperiodic systems but the formalisms are different for the two cases.

The exact $f(\alpha)$ spectrum is obtained for the six-cycle wave function of the Fibonacci model  \cite{KB,FKT}.

\subsection{Multifractal anaysis  of an  energy spectrum}
In order to characterize scaling of an energy spectrum we define
appropriate scaling indices and the entropy function. 
At $\ell$th level of rational approximations, we have $N=2F_\ell$ bands
and their widths are denoted by $\Delta_i$ ($i=1,\cdots,N$). 
Let us define a scaling index for $\Delta_i$ by 
\begin{equation}
 \epsilon_i=-\frac{1}{\ell}\ln{\Delta_i}.
\end{equation}
We also define an entropy functilon by
\begin{equation}
 S(\epsilon )= \lim_{\ell \rightarrow\infty} \frac{1}{\ell}\ln{\Omega_\ell (\epsilon)}, 
 \label{entropy}
\end{equation}
where $\Omega_\ell (\epsilon) d\epsilon$  is the number of bands whose scaling
index lies between $\epsilon$ and $\epsilon+d\epsilon$. 
As in the formalism of statistical mechanics, we
introduce a ``partition function'' and a ``free energy'' which are
defined by
\begin{equation}
 Z_{\ell}(\beta)=\sum_{i=1}^N\Delta^\beta_i,
\end{equation}
and 
\begin{equation}
 F(\beta) = \lim_{\ell \rightarrow\infty}\frac{1}{\ell}\ln{Z_{\ell}(\beta)}.
\end{equation}
Once the free energy is calculated, the enetropy function is obtained by
the Legendre transformation: 
\begin{align}
 &S(\epsilon)=F(\beta)+\beta\epsilon,
 &\epsilon=-\frac{dF(\beta)}{d\beta}.
\end{align}
Thus, by changing $\beta$, one can pick a value of $\epsilon$, and the
corresponding $S(\epsilon)$ is calculated. 

The index $\epsilon$ which represnets scalilng of the Lebesgue measure
of the energy spectrum can actually be related to singular behavior
of the density of states. At $\ell$th level of approximation, each
band carries the same number of states,
\begin{equation}
 p_i=\frac{1}{2F_{\ell}}\sim \frac{1}{\tau^\ell}.
\end{equation}
 where $\tau$ is the golden mean: $(\sqrt 5 + 1)/2$. The total number of states is normalized to unity, i.e., $\sum_ip_i=1$.
An index $\alpha_i$ which represents singular behavior of the
density of states is defined by 
\begin{equation}
 p_i\sim\Delta^{\alpha_i}_i.
\end{equation}
Since $p_i=1/N=\tau^{-\ell}$ and $\Delta_i\sim e^{-\ell \epsilon}$, we have a relation:
\begin{equation}
 \alpha\epsilon=\ln{\tau}.
\end{equation}
The spectrum of singularity introduced by Halsey {\rm et al.}
\cite{Halsey}  is given by 
\begin{equation}
 \Omega'(\epsilon, \alpha)=\langle \Delta\rangle^{-f(\alpha)}, 
\end{equation}
where $\Omega'(\epsilon,\alpha)d\epsilon d\alpha$ is the number of bands
whose scaling index lies between $\epsilon$ and $\epsilon+d\epsilon$,
and $\alpha$ and $\alpha+d\alpha$, and $\langle\Delta\rangle$  is a
representative value of $\Delta$, $\langle \Delta\rangle=\exp{(-\ell \epsilon)}$. Then  $f(\alpha)$ is related to the entropy function $S(\epsilon)$ by
\begin{equation}
 f(\alpha)=\frac{S(\epsilon)}{\epsilon}.
\end{equation}
The point here is that wether the limit $\ell \rightarrow \infty$ in Eq. (\ref{entropy}) exists or not. Only when it exists the energy spectrum is multifractal.

\begin{figure}[t]
 \includegraphics[width=65mm]{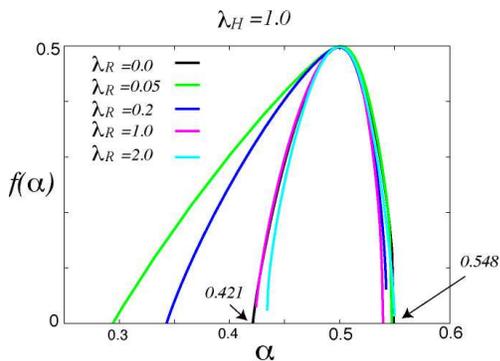} 
\caption{\label{alpha-spectrum} (Color online) $f(\alpha)$ for energy spectrum on the self-dual line.  At $\lambda_R=0$ it corresponds to that of Harper equation.}
\end{figure}

We apply the method above to the energy spectrum  on the self-dual
line, $\lambda_H=1$. It is possible to obtain $f(\alpha)$ numerically by extrapolation of large $\ell$, and some of the results are shown in Fig.~\ref{alpha-spectrum}. These $f(\alpha)$'s 
exhibit smooth curves in intervals [$\alpha_{\text{min}},\alpha_{\text{max}}$]. 
This is an evidence that the energy  spectrum is singular
continuous and we conclude that all the wave functions on the self-dual line is critical.

\subsection{Multifractal analysis of wave functions}
Take a period of the  lattice which has a Fibonacci number of lattice points. Normalize the length of the period by unity by putting  the lattice spacing as
\be
l_i=l=\frac{1}{F_{\ell}}\sim\frac{1}{\tau^\ell}.
\ee
The scaling index for $l_i$, which is given by
$l_i=\exp(-\ell\epsilon_i)$, is a constant: 
\be
\epsilon=-\frac{1}{\ell}\ln{l}\sim\ln{\tau}.
\ee
We define a  probability measure from a wave function by
\be
p_i = |\psi_{i}|^2,
\ee
which is normalized as
\begin{equation}
 \sum_{i=1}^{F_{\ell}} p_i = \sum_{i = 1}^{F_{\ell}} |\psi_{i}|^2=1.
\end{equation}
The scaling index for $p_i$ is defined by
\begin{equation}
 p_i=l_i^\alpha,
\end{equation}
and the entropy function for $\alpha$ is defined by
\begin{equation}
 S(\alpha) = \lim_{\ell \rightarrow \infty} \frac{1}{\ell}\ln{\Omega_\ell(\alpha)}, 
\label{sa} 
\end{equation}
where $\Omega_\ell (\alpha)d\alpha$ is the number of sites which have an index
between $\alpha$ and $\alpha+d\alpha$. 

As in the previous case of the energy spectrum we introduce a partition function:
\begin{equation}
 Z_{\ell}(q)=\sum_{i}p^q_i, 
\end{equation}
and a free energy:
\begin{equation}
 G(q)=\lim_{\ell \rightarrow\infty}\frac{1}{\ell}\ln{Z_{\ell}(q)}.
\end{equation}
The entropy function is given by the Legendre transformation:
\be
 S(\alpha) &=& G(q)+q \alpha \epsilon, \nonumber \\
\alpha &=& -\frac{1}{\epsilon}\frac{dG(q)}{dq}.
\ee
Function $f(\alpha)$ in this case is defined by 
\begin{equation}
 \Omega(\alpha)=l^{-f(\alpha)}, 
\end{equation}
and is  related to the entropy functiton as
\begin{equation}
 f(\alpha)=\frac{S(\alpha)}{\epsilon}.
\end{equation}

We calculate $f(\alpha)$ for finite Fibonacci indices $\ell$ and extrapolate them to $\ell \rightarrow\infty$. Actually only a part of
$f(\alpha)$ is required to distinguish localized, extended, and critical
wave functions. 
For extended wave functions, the probability measure scales as $p_i\sim
1/N=l$, thus we have $\alpha_{\text{min}}=1$. For localized wave functions, on
the other hand, $p_i$ is finite ($\alpha=0$) only on a finite number of
sites and is exponentially small ($\alpha=\infty$) on the other sites, 
thus we have $f(\alpha)$ which is defined at $\alpha=0$ [$f(0)=0$] and $\alpha=\infty$
[$f(\infty)=1$]. For critical wave functions, $\alpha$ has a
distribution, i.e., $f(\alpha)$ is a smooth function defined on a finite
interval [$\alpha_{\text{min}}$, $\alpha_{\text{max}}$]\cite{HiramotoKohmoto}. 
Thus we need to calculate $\alpha_{\text{min}}$ which corresponds to the
largest $|\psi_j|^2$ to distinguish extended, localized and
critical wave functions. Namely, 
\begin{align}
 &\alpha_{min}=1
\hspace{10mm}\text{for an extended wave function}, \nonumber \\
&\alpha_{min}\ne 0,1
\qquad\text{for a critical wave function}, \nonumber \\
&\alpha_{min}=0
\hspace{10mm}
\text{for a localized wave function}.
\end{align}

\begin{figure}[t]
\begin{center}
 \includegraphics[width=70mm]{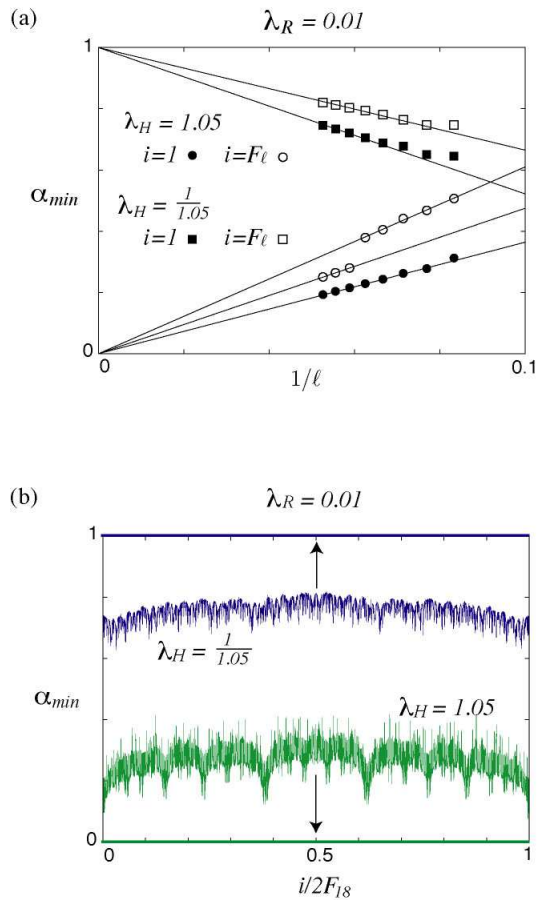} 
 \end{center}
\caption{\label{alpha} (Color online)
(a)  Plots of $\alpha_{min}$ vs. $1/\ell$ 
 for the wave functions of the lowest band  ($i = 1$) and the center band ($i = F_\ell$).
(b) Plots of $\alpha_{min}$ for the wave function of each band. The number of wave functions
 is $2F_{18}$ ($F_{18} = 4181$). }
\end{figure}

At $\ell$th approximation of the Cantor set sectrum, 
there are $N=2F_\ell$ energy bands. We obtain a wave function with $(k_x, k_y) = (0, 0)$ at each band.
 Examples of determining $\alpha_{min}$ for wave functions in Phase I and Phase IV are shown in Fig.~\ref{alpha} (a). The wave functions of the lowest band are denoted as $i = 1$. For this wave function $\alpha_{min}$ extrapolates to $0$ if $(\lambda_H, \lambda_R) = (1.05, 0.01)$, and to $1$ if $(\lambda_H, \lambda_R) = (1/1.05, 0.01)$ (see Fig. \ref{alpha} (a)). In this manner it is determined that $\alpha_{min} = 0$ for all the wave functions if $(\lambda_H, \lambda_R) = ( 1.05, 0.01)$, and  $\alpha_{min} = 1$ for all the wave functions if $(\lambda_H, \lambda_R) = (1/ 1.05, 0.01)$ (see Fig. \ref{alpha} (b)).
The wave functions at the center of the spectrum ($i = F_{\ell}$) give the same results for $\alpha_{min}$.
In this way the multifractal analysis of wave function also shows that all the wave functions are extended in Phase I and all the wave functions are localized in Phase IV.

\begin{figure}[t]
 \includegraphics[width=70mm]{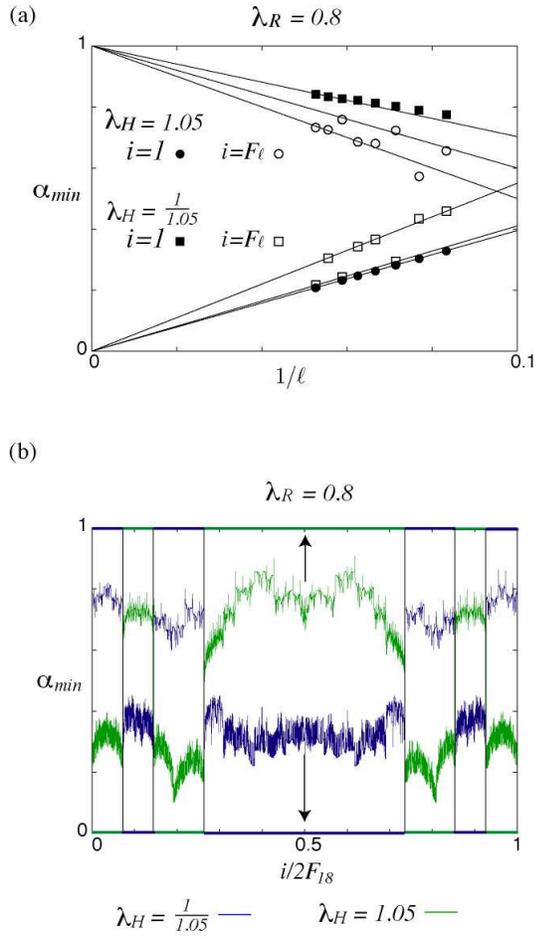}
\caption{\label{alpha-min} (Color online)
(a) Plot of $\alpha_{min}$ vs. $1/\ell$ for the wave functions
of the lowest band ($i = 1$) and the center band ($i = F_\ell$).
(b) Plots of $\alpha_{min}$ for the wave function of each band. 
The number of wave functions is $2F_{18}$ ($F_{18} = 4181$). }
\end{figure}
Examples of the results for Phase II and Phase III are shown in Fig.~\ref{alpha-min}. For the wave function of the lowest band ($i = 1$)  $\alpha_{min}$ extrapolates to $0$ if $(\lambda_H, \lambda_R) = ( 1.05, 0.8)$, and to $1$ if $(\lambda_H, \lambda_R) = ( 1/1.05, 0.8)$ (see Fig. \ref{alpha-min} (a)). On the other hand, $\alpha_{min}$ for the wave functions at the center of the spectrum ($i = F_{\ell}$) $\alpha_{min}$ extrapolates to $1$ if $(\lambda_H, \lambda_R) = ( 1.05, 0.8)$, and to $0$ if $(\lambda_H, \lambda_R) = ( 1/1.05, 0.8)$ (see Fig. ~\ref{alpha-min} (a)). This suggests that there exist a number of mobility edges in the energy spectrum. The existence of mobility edges is numerically shown using this multifractal analysis of wave function, and an example of the data are plotted in Fig.~\ref{alpha-min} (b).

The duality discussed in Sec. \ref{hamiltonian} B dictates that, if a wave function with
$(\lambda_H,\lambda_R)=(\alpha,\beta)$ is extended, 
the corresponding wave function with
$(\lambda_H,\lambda_R)=(1/\alpha, \beta)$
is localized, and vise versa. This behavior is clearly seen in Fig.~\ref{alpha-min} (b).

\section{Wave Functions}
\label{WF}
In this section examples of the wave functions in pahases I, II, III, and IV; and on the self-dual line are displayed.

\begin{figure}[t]
\includegraphics[width=55mm]{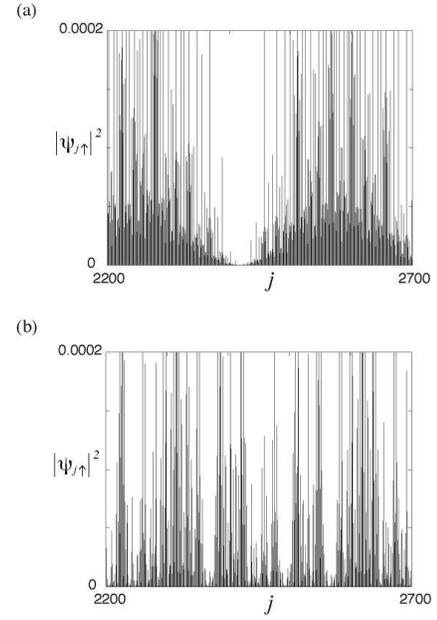}
\caption{\label{wf-I}  Square moduli of wave functions in Phase I. $(\lambda_H, \lambda_R) = (1/1.05,  0.01), \ell =19, F_{\ell}=6765$. 
(a) At the edge of the energy spectrum,  (b) at
 the center of the energy spectrum.}
\end{figure}

\begin{figure}
\includegraphics[width=55mm]{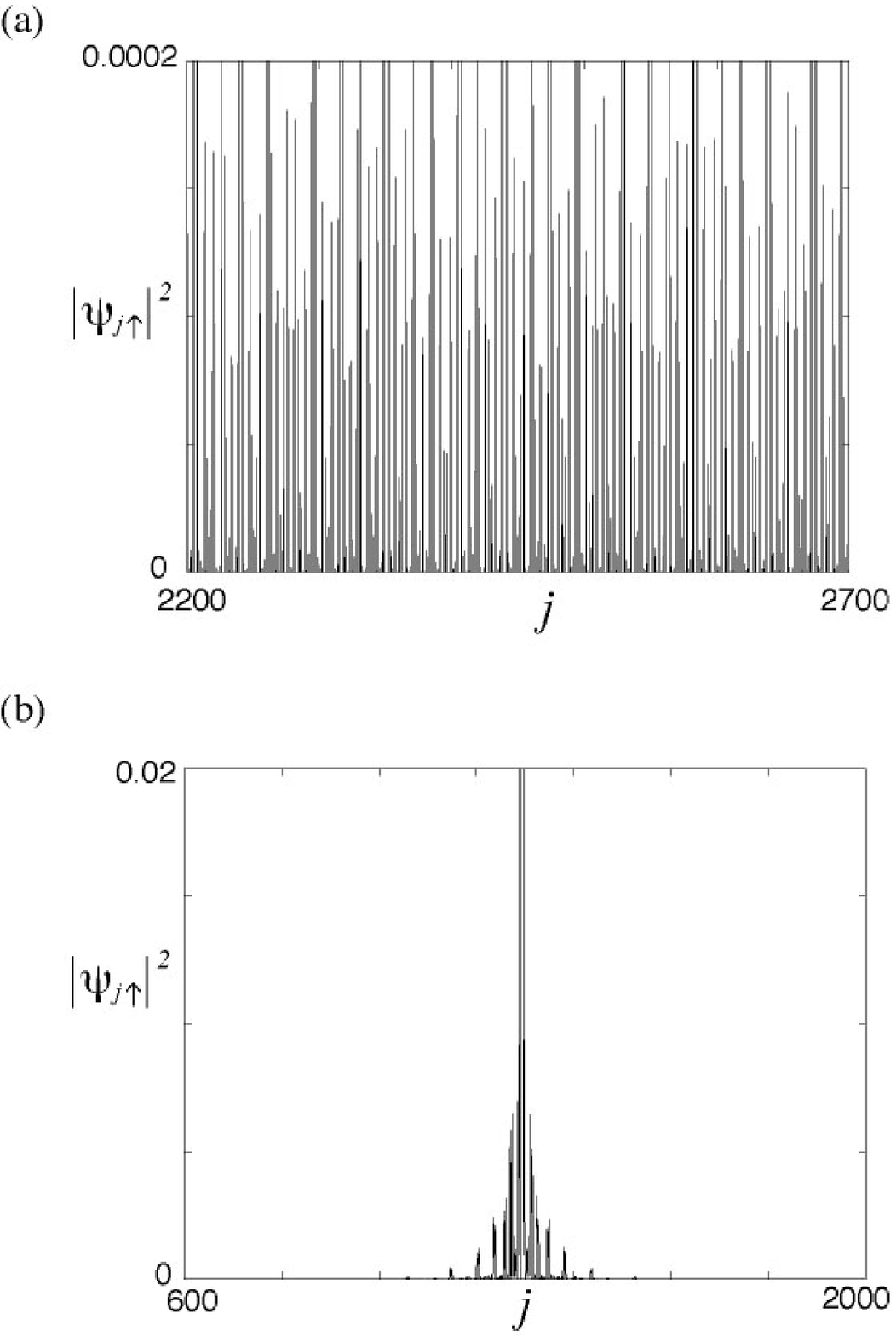}
\caption{\label{wf-II}  Square moduli of wave functions in Phase II. 
$(\lambda_H,\lambda_R)=(1/1.05, 0.8)$, $\ell=19$, $F_{\ell}=6765$
  (a) At the edge of the energy spectrum,  (b) at
 the center of the energy spectrum. }
\end{figure}

\begin{figure}
\includegraphics[width=55mm]{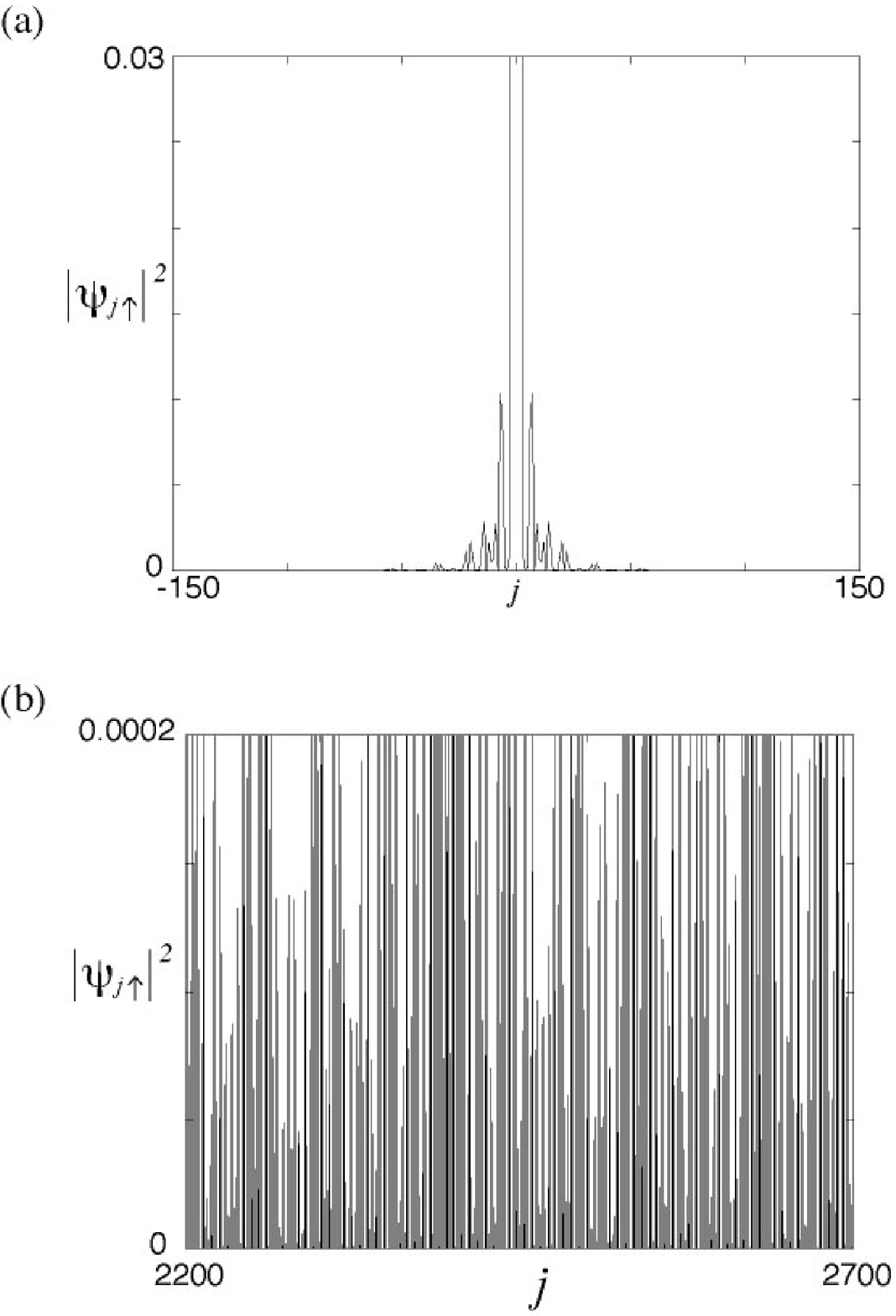}
\caption{\label{wf-III}Square moduli of wave functions in Phase III. $(\lambda_H,\lambda_R)=(1.05, 0.8), \ell=19, F_{\ell}=6765$.  
 (a) At the edge of the energy spectrum,  (b) at
 the center of the energy spectrum. }
\end{figure}

\begin{figure}
\includegraphics[width=55mm]{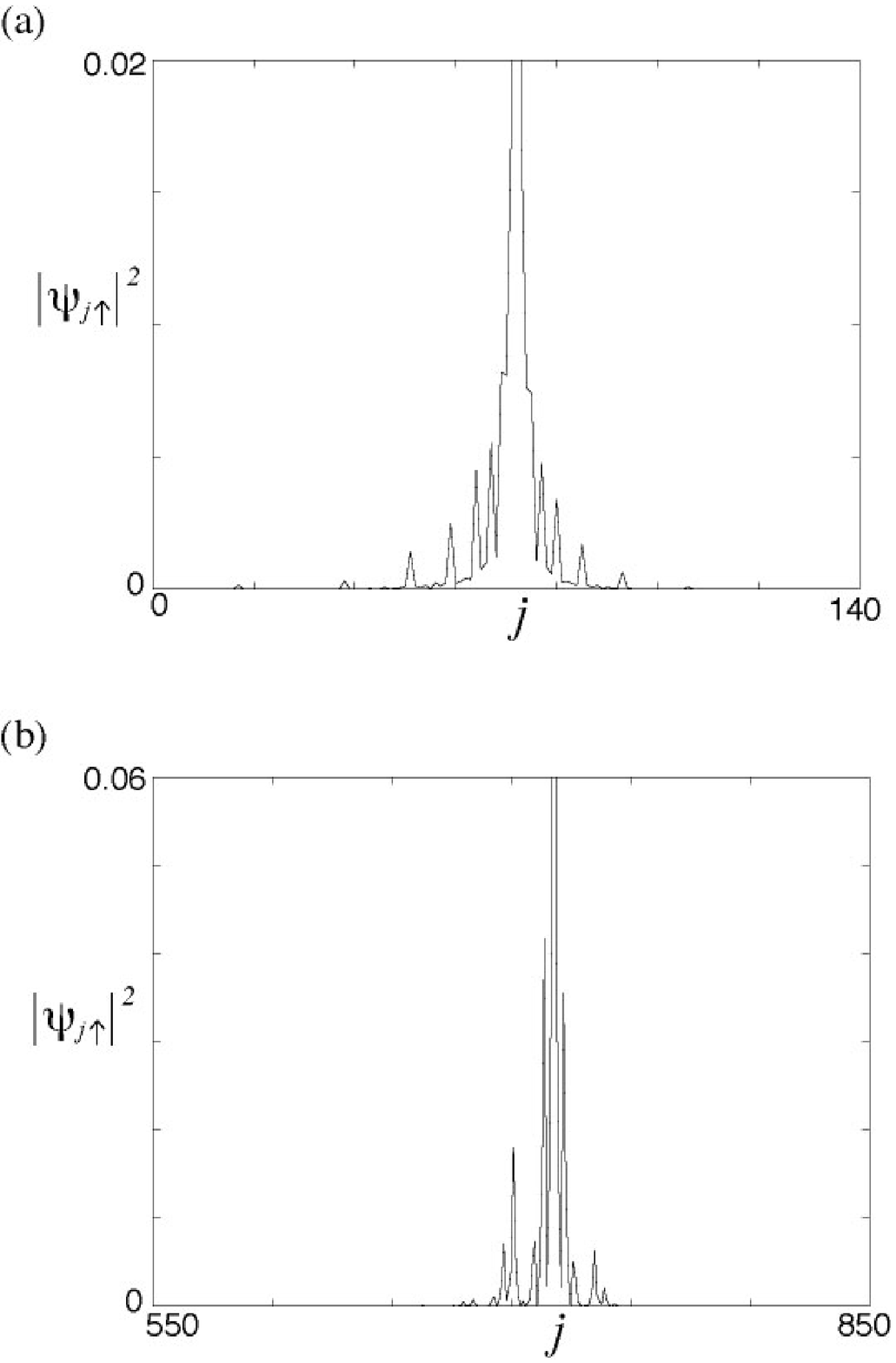}
\caption{\label{wf-IV}
Square moduli of wave functions  in Phase IV. $(\lambda_H,\lambda_R)=(1.05,0.01),  \ell = 19, F_{19} = 6765$.
  (a) At the edge of the energy spectrum,   (b) at the center of the energy spectrum. }
\end{figure}

\begin{figure}[t]
\includegraphics[width=55mm]{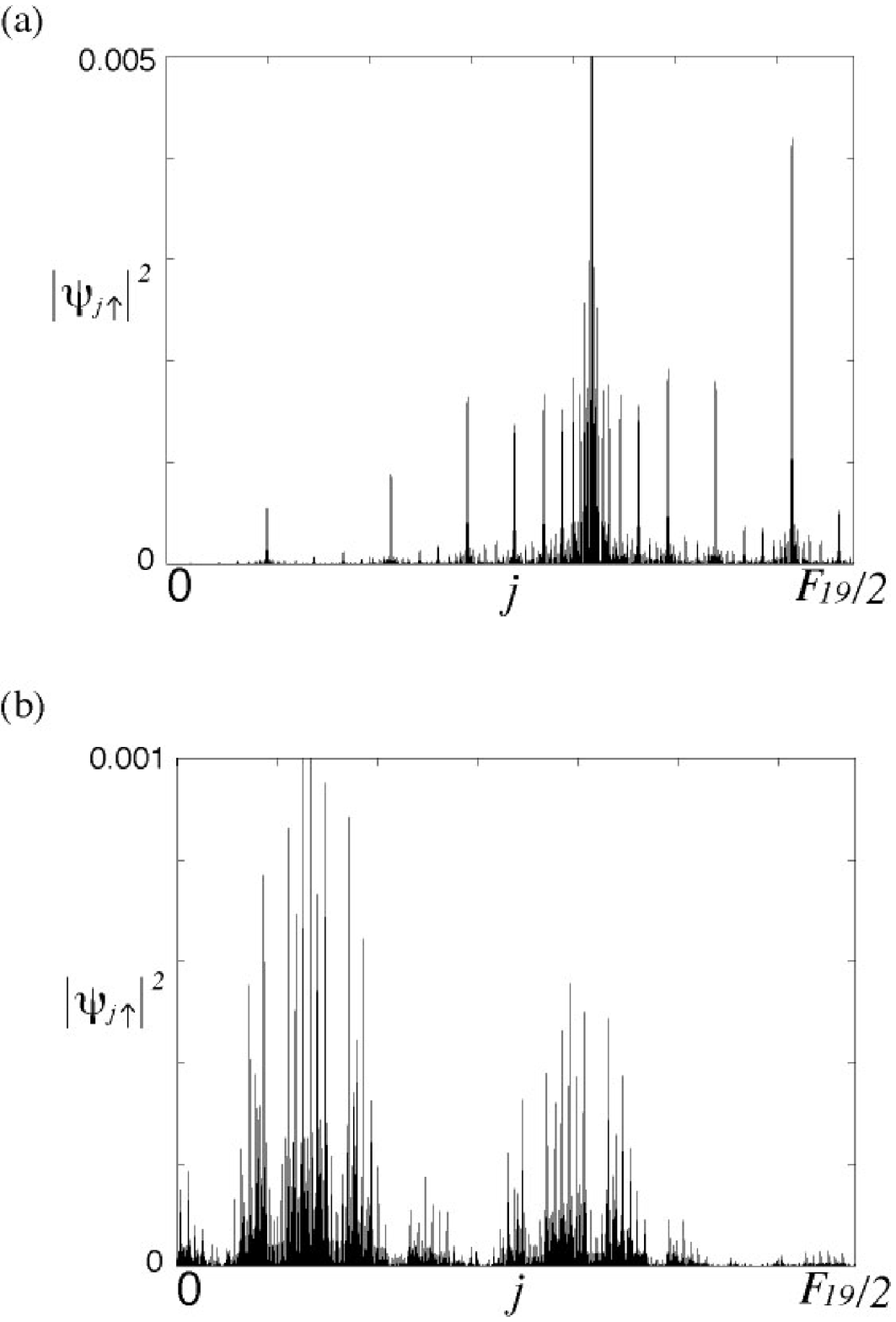}
\caption{\label{wf-critical} Square moduli of wave functions on the self-dula line, $(\lambda_H,\lambda_R)=(1, 0.2), \ell=19, F_{\ell}=6765$. (a) At the edge of the energy spectrum, (b) at the center of the energy spectrum. }
\end{figure}

Example of wave functions in Phase I are shown in Fig.~\ref{wf-I}.
All the  wave functions are extended in Phase I. 
Examples of wave functions in Phase II are shown in Fig.~\ref{wf-II}. 
We see that the wave function at the edge
of the energy spectrum is extended and the wave function at the center of the
energy spectrum is localized. 
Examples of wave functions in Phase III are shown in Fig.~\ref{wf-III}. 
We see that the wave function at the edge
of the energy spectrum is localized and the wave function at the center of the
energy spectrum is extended. 
Figure.~\ref{wf-IV} is  examples of wave functions in Phase IV in which all the  wave functions are
localized. 
Examples of critical wave functions on the self-dual line are shown in Fig.~ \ref{wf-critical}. 
{\acknowledgements}  
We are grateful to C. Wexler for useful discussions.

\clearpage


\end{document}